\documentclass[aps,showpacs,prl,twocolumn]{revtex4}
\usepackage{amsmath,amssymb}
\usepackage{graphics,epsfig}
\usepackage{graphicx}
\usepackage{color}

\def\build#1_#2^#3{\mathrel{\mathop{\kern 0pt#1}\limits_{#2}^{#3}}}

 \newcommand{\vct}[1]{{\mbox {\boldmath $#1$}}}

\begin{document}

\title{Lagrangian dynamics and statistical geometric structure of turbulence}

\author{L. Chevillard and C. Meneveau}
\affiliation{Department of Mechanical Engineering and Center for
Environmental and Applied Fluid Mechanics, The Johns Hopkins
University, 3400 N. Charles Street, Baltimore, MD 21218, USA}

\begin{abstract}
The local statistical and geometric structure of three-dimensional
turbulent flow can be described by properties of the velocity
gradient tensor. A stochastic model is developed for the
Lagrangian time evolution of this tensor, in which the exact
nonlinear self-stretching term accounts for the development of
well-known non-Gaussian statistics and geometric alignment trends.
The non-local pressure and viscous effects are accounted for by a
closure that models the material deformation history of fluid
elements.  The resulting stochastic system reproduces many
statistical and geometric trends observed in numerical and
experimental 3D turbulent flows, including anomalous
relative scaling.

\end{abstract}


\maketitle

Fully developed turbulent flows are omnipresent in the natural and
man-made environment. Development of deeper understanding of
fundamental properties of turbulence is needed for progress in a
number of important fields such as  meteorology, combustion,  and
astrophysics. Despite the highly complex nature of inherently
three-dimensional velocity fluctuations, turbulent flows exhibit
universal statistical properties. An example is the $k^{-5/3}$-law
of Kolmogorov \cite{K41}.  Another example is the ubiquity of
intermittency of longitudinal and transverse Eulerian velocity
increments between two points \cite{Fri95}. Moreover, probability
density functions (PDFs) of velocity increments  change with the
length-scale between the points. Starting from  an almost Gaussian
density at large scale $L$ (i.e. the integral length scale), these
PDFs undergo a continuous deformation in the inertial range to
finish in a highly skewed and non Gaussian PDF near the viscous
scale of turbulence \cite{Fri95,CheCas06}. The latter is,
equivalently, also true for the velocity gradients. Recently a
simple two-equation dynamical system was derived \cite{LiMen0506}
that reproduces the formation of intermittent tails in the PDFs.

While much attention has been devoted to the statistics and
anomalous scaling of longitudinal and transverse velocity
increments, there has been growing interest  (see e.g. \cite{Zeff}) in
the properties of the full velocity gradient tensor $A_{ij} =
\partial_ju_i$. $A_{ij}$ characterizes variations of all
velocity components, in all directions. Such
additional information is required (but unavailable) to model
pressure effects in the system of Refs. \cite{LiMen0506} and thus
to allow reproducing stationary statistics. Empirically it has
also become apparent that $\textbf{A}$ displays a number of
interesting and possibly universal geometric features. For
example, the vorticity vector (related to the antisymmetric part
of  $\textbf{A}$) is preferentially aligned \cite{GeoVortExpNum}
with the eigenvector of the intermediate eigenvalue of the
strain-rate tensor $\textbf{S}=(\textbf{A}+\textbf{A}^T)/2$, where
$T$ stands for transpose. Moreover, the preferred state of the
local deformation is axisymmetric expansion, corresponding to two
positive and one negative eigenvalues of $\textbf{S}$. These
geometric trends have been repeatedly observed in experimental and
numerical experiments \cite{GeoVortExpNum}, both at the viscous
scale as well as in the inertial range, for a variety of different
flows. These trends can be readily understood from the nonlinear
self-stretching \cite{Viel84,Can91} that occurs during the
Lagrangian evolution of $\textbf{A}$. However, the resulting
so-called Restricted Euler (RE) dynamics, obtained by neglecting
viscous diffusion and the non-local anisotropic effects of
pressure, display unphysical finite-time singularities. These
are due to the absence of regularization properties
of the neglected viscous and pressure gradient terms. Prior models
that seek to regularize the RE dynamics include a stochastic model
in which the nonlinear term is modified to yield, by construction,
log-normal statistics of the dissipation  \cite{GirPop90_1}, a
linear damping model for the viscous term \cite{Martin}, and the
tetrad model \cite{ChePum99} in which the material deformation
history is used to model the unclosed pressure Hessian term.
Material deformation is also tracked in the viscous diffusion
closure in Ref. \cite{JeoGir}. While each of these models add
useful features, a model that has no singularities and leads to
stationary statistics, without  tuning the nonlinear term explicitly
to impose log-normal dissipation statistics, is still lacking.
The aim of this Letter is to introduce such a model and to document
its properties.

The Lagrangian evolution of $A_{ij}$ is governed by the gradient
of the incompressible Navier-Stokes equations:
\begin{equation}\label{eq:NS}
\frac{dA_{ij}}{dt} = -A_{ik}A_{kj} - \frac{\partial ^2 p}{\partial
x_i \partial x_j } + \nu \frac{\partial ^2 A_{ij}}{\partial x_m
\partial x_m}\mbox{ ,}
\end{equation}
where $\nu$ is the kinematic viscosity, $p$ is the pressure
divided by density, and $d/dt$ the Lagrangian (material)
derivative. $A_{ii}=0$ at all times. The last two terms in Eq.
(\ref{eq:NS}) are unclosed. If the pressure Hessian
$\partial^2_{ij}p$ is assumed to be an isotropic tensor, its trace
can be expressed in terms of an invariant of ${\bf A}$ which
yields, together with neglect of the viscous term, to the
above-mentioned, closed, RE system \cite{Can91}. Yet, it is
well-known that it is unphysical to assume that $\partial^2_{ij}p$
is isotropic, given the complex anisotropic effects of pressure
gradient.

\begin{figure}[t]
\center{\epsfig{file=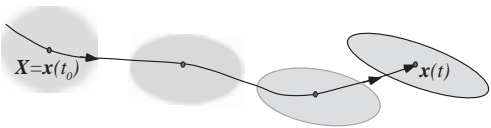,width=6.5cm}}
\caption{\label{fig:Sketch} Sketch of how an initially uncertain
(and thus modeled as isotropic) material element is mapped onto
the present position ${\bf x}$ at time $t$, reflecting recent
deformations.}
\end{figure}

If instead we focus on changes of local pressure with changes of
past fluid particle locations (${\bf X}$) at some early time in
the Lagrangian history (i.e. focus on  the Lagrangian pressure
Hessian $P_{mn} \equiv \partial^2 p/\partial X_m \partial X_n$,
where $p$ is  evaluated at present time $t$ but as function of
initial positions), the assumption of isotropy is better
justified. This is based on the idea that any causal relationship
between the initial time and the present has been lost due to the
stochastic nature of turbulent dispersion. The sketch in Fig.
\ref{fig:Sketch} is meant to describe how an initially uncertain
(and thus modelled as isotropic) material shape is mapped onto the
present location with a deformed shape that mirrors the recent
local deformations due to the velocity gradient history.  The
notation is as follows: ${\bf x}(t)$ denotes the present position
of interest, at time $t$. $\mathcal M_{t_0,t}:{\bf X}\mapsto {\bf
x}$ is the Lagrangian path map \cite{Con01} which gives the
Eulerian position  ${\bf x}$ at time $t$ of a fluid particle
initially located  at the position ${\bf X}$ at time $t_0$.  By
virtue of incompressibility, this map is invertible and its
Jacobian (the deformation gradient tensor), $D_{ij} = \partial
x_i/\partial X_j$ has  determinant $\mbox{det}(\textbf{D})=1$ at
any time \cite{MonYag}. We denote its inverse by $D_{ij}^{-1} =
\partial X_i/\partial x_j$. The tensor $C_{ij}=D_{ik}D_{jk}$ is
called the Cauchy-Green tensor which has been studied in turbulent
flows numerically and experimentally \cite{GirPopJFM,TsinoLag}.

The relationship between the Eulerian and Lagrangian pressure
Hessian is obtained by applying twice the change of variables
$\partial /\partial x_j = (\partial X_m/\partial x_j) \partial
/\partial X_m$, and neglecting $\partial (\partial X_m/\partial
x_j)/\partial x_i$ (i.e. neglecting spatial variations of
$\bf{D}^{-1}$ \cite{Con01}). Then, the main closure hypothesis is
that the Lagrangian pressure Hessian, $P_{mn}$, is isotropic (i.e.
$P_{mn} = P_{kk} \delta_{mn}/3$, where $\delta_{mn}$ is the
Kronecker tensor), when the time-delay $t-t_0$ is long enough to
justify loss of information. The pressure Hessian can then be
rewritten according to
\begin{equation}\label{eq:ChangVar}
\frac{\partial ^2 p}{\partial x_i \partial x_j } \approx
\frac{\partial X_m}{\partial x_i}\frac{\partial X_n}{\partial
x_j}\frac{\partial ^2 p}{\partial X_m \partial X_n }~=
C^{-1}_{ij}\frac{1}{3}P_{kk},
\end{equation}
which could be regarded as a reinterpretation of the  ``tetrad
model'' \cite{ChePum99}.

The dynamics of  $\textbf{D}$ are determined by $d\textbf{D}(t)/dt
= \textbf{A}(t)\textbf{D}(t)$. Starting at some initial time from
$D_{ij}(t_0) = \delta_{ij}$, the general form of $\textbf{D}$  can
be written formally using the time-ordered exponential function
(${\rm exp}_{\mathcal T}$), i.e. $\textbf{D}(t) = {\rm
exp}_{\mathcal T}\left[ \int_{t_0}^t ds \textbf{A}(s)\right]$
\cite{FalGaw01}.

To determine $P_{kk}$, we follow Ref. \cite{ChePum99} and use the
Poisson equation $\nabla^2p=-A_{nm}A_{mn}=C^{-1}_{qq}P_{kk}/3$,
from which $P_{kk}$ can be solved, leading to \cite{ChePum99}
\begin{equation}\label{eq:ModTet}
\frac{\partial ^2 p}{\partial x_i \partial x_j
}=-\frac{C_{ij}^{-1}}{C_{qq}^{-1}}A_{nm}A_{mn}\mbox{ .}
\end{equation}
A similar approach can be applied \cite{JeoGir} to the viscous
term, expressing the Laplacian of $A_{ij}$ in the Lagrangian
frame, as in Eq. (\ref{eq:ChangVar}). The resulting Lagrangian
Hessian of $A_{ij}$ is modeled by a classical linear damping term,
namely $\nu {\partial^2 A_{ij}}/{\partial X_p \partial X_q }
\approx -\delta_{pq} A_{ij}/(3T) $. The relaxation time-scale $T$ is chosen to be on the order of the
integral time-scale. This can be justified by recognizing
that the distance travelled by a viscous eddy during a viscous turn-over or decorrelation time,
advected by the rms turbulence velocity $u'$,
scales like the Taylor microscale, $\lambda$.
Assuming therefore that $\lambda$ is the appropriate Lagrangian
decorrelation length-scale of $A_{ij}$, it follows
that $\nu/(\partial X)^2 \sim \nu/\lambda^2 \sim 1/T$.
Finally, the model reads
\begin{equation}\label{eq:ModLap}
\nu\frac{\partial ^2 A_{ij}}{\partial x_m \partial x_m
}=-\frac{1}{T}\frac{C_{mm}^{-1}}{3}A_{ij}\mbox{ ,}
\end{equation}
and is reminiscent of mapping closures \cite{Kra90}.

Replacing the pressure Hessian and the viscous term in Eq.
(\ref{eq:NS}) by the modeled terms Eqs. (\ref{eq:ModTet}) and
(\ref{eq:ModLap}), one can show numerically that the finite-time
divergence induced by the quadratic term is regularized, and each
component of $A_{ij}$ tends to zero at long times. Next, to
generate stationary statistics a stochastic forcing term can be
added. The resulting system, however, is not stationary since it
depends upon the evolving tensors $\textbf{D}$ and $\textbf{C}$
whose time evolutions reflect the non-stationary nature of
turbulent dispersion. For example, on average the largest (resp.
smallest) eigenvalue of $\textbf{C}$ undergoes exponential growth
(resp. decrease) in time, whereas the intermediate one remains
approximatively constant \cite{MonYag,GirPopJFM,TsinoLag}.  We
remark that in the tetrad model \cite{ChePum99} this feature is
exploited to keep track of changing length scale. Our aim here is
to develop a statistically stationary description of the velocity
gradient at a fixed scale (e.g. viscous scale).

The crucial step of the proposed model is to replace the actual
slow decorrelation along the Lagrangian trajectory and the total
deformation history $[t_0,t]$  with a perfect correlation of
$A_{ij}$ during a time scale $\tau$ (which is thought to be of the order
of the Kolmogorov time-scale $\sqrt{\nu/\epsilon}$, where $\epsilon$ is the
dissipation rate).  Correlations for time-delays longer than $\tau$
are neglected. It follows, using the time-ordered exponential property, that $\textbf{D}(t)
=\textbf{D}(t-\tau)\textbf{D}_{\tau}(t)$, where
$\textbf{D}_{\tau}(t)\approx e^{\tau\textbf{A}(t)}$. Furthermore,
we neglect the prior deformation history. Accordingly, we may
define a ``stationary Cauchy-Green tensor''
\begin{equation}\label{eq:CauchyGreen}
\textbf{C}_{\tau}(t) = \textbf{D}_{\tau}(t)\textbf{D}_{\tau}^T(t)
= e^{\tau \textbf{A}}e^{\tau \textbf{A}^T}\mbox{ .}
\end{equation}
When $\tau$ decreases (i.e. the Reynolds number $\mathcal R_e$
increases), at fixed ${\bf A}$ the restitution strength of the pressure Hessian model
decreases  ($\tau=0$ corresponds to an isotropic pressure Hessian as
in the singular RE system). Without loss of generality, henceforth all variables will be
scaled with the time-scale $T$, i.e.  $t/T \to t$ and
$A_{ij} T\to A_{ij} $.  Combining Eqs. (\ref{eq:NS}),
(\ref{eq:ModTet}), (\ref{eq:ModLap}), (\ref{eq:CauchyGreen}) and a
forcing term, and defining the parameter $\Gamma \equiv \tau/T$ ($\sim \mathcal
R_e^{-1/2}$), the following stochastic differential equation is
finally obtained:
\begin{equation}\label{eq:ourmodel}
d\textbf{A} = \left( -\textbf{A}^2+
\frac{\mbox{Tr}(\textbf{A}^2)}{\mbox{Tr}(\textbf{C}_{\Gamma}^{-1})}
\textbf{C}_{\Gamma}^{-1}
-\frac{\mbox{Tr}(\textbf{C}_{\Gamma}^{-1})}{3} \textbf{A}\right)dt
+d\textbf{W}\mbox{ .}
\end{equation}
\begin{figure}[t]
\center{\epsfig{file=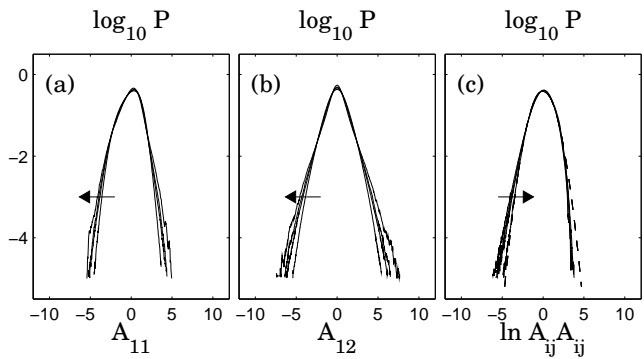,width=8.5cm}}
\caption{\label{fig:IncVar} (a) and (b):  PDFs of longitudinal and
transverse components of the velocity gradient tensor $\textbf{A}$
(normalized by its root-mean-square), obtained from
time-integration of Eq. (\ref{eq:ourmodel}) for $\Gamma=0.2, 0.1,
0.08$ and 0.06. (c) PDF of $\ln A_{ij}A_{ij}$, for the same values
of $\Gamma$. A Gaussian PDF of unit variance (dashed line) is also
shown. Arrows direction indicates decreasing $\Gamma$ (increasing
$\mathcal R_e$). }
\end{figure}
The tensorial noise $d\textbf{W}$ represents physical effects that
have been neglected, such as action of larger-scale, and
neighboring, eddies.  For simplicity, we assume $d\textbf{W}$ is
Gaussian and white in time. In the assumed units of time, we choose
$d\textbf{W} = \textbf{G}\sqrt{2dt}$, where $\textbf{G}$ is a
tensorial Gaussian, delta-correlated noise. Its covariance matrix
should be consistent with an isotropic, homogeneous, and traceless
tensorial field, namely $\langle G_{ij}G_{kl}\rangle =
2\delta_{ik}\delta_{jl}-\frac{1}{2}\delta_{ij}\delta_{kl}-
\frac{1}{2}\delta_{il}\delta_{jk}$  \cite{PopBook}. When $dW_{ij}
=0$, numerical tests show that the finite-time divergence is
regularized for any initial condition.

The stochastic differential equation (\ref{eq:ourmodel}) is solved
numerically using four different values for $\Gamma$: 0.2, 0.1,
0.08 and 0.06. A second-order weak predicator-corrector scheme
\cite{KloPla92} is used, with time steps $dt = 10^{-2}$ ($dt =
10^{-3}$ is used for $\Gamma=0.06$). Integration times of order
$10^5 ~T$'s are used. Time-series of each component of
$\textbf{A}$ indicate stationary behavior. In Figs.
\ref{fig:IncVar}(a-b) we show the PDFs of longitudinal ($A_{11}$)
and transverse ($A_{12}$) components for various $\Gamma$ values
(here and below, all statistics are improved by averaging over all
available longitudinal and transverse directions, respectively).
When $\Gamma$ decreases, velocity gradient PDFs develop slightly longer tails.
Also, the longitudinal components are negatively skewed.

It has been observed in numerical simulations \cite{PopChe90} that
the pseudo-dissipation $A_{ij}A_{ij}$ is close to lognormal for
any Reynolds number (as obtained in the stationary diffusion
process \cite{GirPop90_1} by specific construction of the
nonlinear term), and one wonders whether lognormality arises in
the present model. Fig. \ref{fig:IncVar}(c) presents the PDF of
the logarithm of the pseudo-dissipation for various values of the
parameter $\Gamma$.  The PDF of $\ln A_{ij}A_{ij}$ from the
model is close (but not exactly equal) to Gaussian. Note that the finiteness of dissipation
implies that $\langle A_{11}^2\rangle /T^2 = \epsilon/(15\nu)$. It follows that $\tau/\sqrt{\nu/\epsilon}$ is
fixed through $\tau^2/(\nu/\epsilon)=15\langle A_{11}^2\rangle \Gamma^2$.

\begin{figure}[t]
\center{\epsfig{file=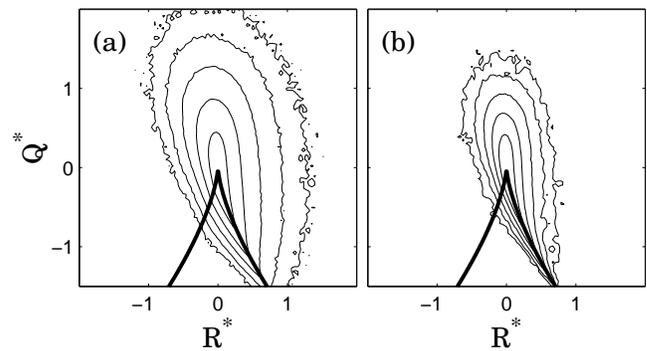,width=8.5cm}} \caption{Contour
plots of the logarithm of the joint probability of the two
invariants of the velocity gradient tensor (see text)
non-dimensionalized by the average strain, i.e. $Q^* = Q/\langle
S_{ij}S_{ij}\rangle$ and $R^* = R/\langle
S_{ij}S_{ij}\rangle^{3/2}$, for (a) $\Gamma= 0.2$ and (b)
$\Gamma=0.08$. Contours are logarithmically
spaced, starting at 1 and separated by factors of 10. The thick
line corresponds to zero discriminant (Vieillefosse line).}
\label{fig:RQVar}
\end{figure}

To further characterize the statistics of  $\textbf{A}$, Fig.
\ref{fig:RQVar} presents  the  joint PDF of two important
invariants of $\textbf{A}$, namely $Q=-\mbox{Tr}(\textbf{A}^2)/2$
and $R=-\mbox{Tr}(\textbf{A}^3)/3$, non-dimensionalized by
$\langle S_{ij}S_{ij} \rangle$. The joint PDF in the RQ-plane
shows the characteristic teardrope shape observed in various
numerical and experimental studies \cite{GeoVortExpNum,ChePum99}
and is consistent with predominance of enstrophy-enstrophy
production (top-left quadrant) and dissipation-dissipation
production (bottom-right quadrant). For decreasing $\Gamma$, the
joint PDF becomes more elongated along the right tail of the
Vieillefosse line, consistent with data at increasing $\mathcal
R_e$ \cite{GeoVortExpNum,ChePum99}.

\begin{figure}[t]
\centerline{\epsfig{file=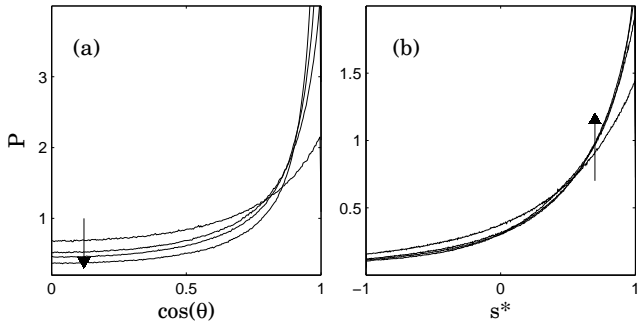,width=8.5cm}}
\caption{\label{fig:GeoVar} Alignment trends and preferred
strain-rate state. In (a) is displayed the PDF of  the cosine of
the angle between vorticity and the intermediate eigenvector of
the strain-rate tensor, showing preferential alignment. In (b),
the PDF of the strain-rate state $s^*$ is presented.}
\end{figure}

Next, the statistics of alignment of the vorticity vector
$\omega_i=\varepsilon_{ijk}A_{kj}$ with
$\textbf{S}$, and the $\textbf{S}$-eigenvalues $\alpha$, $\beta$ and $\gamma$
 are quantified.  In Fig. \ref{fig:GeoVar}(a) the
PDF of $\cos(\theta)$, where $\theta$ is the angle between
${\vct{\omega}}$ and the $\textbf{S}$-eigenvector corresponding to
its intermediate eigenvalue, is shown. Clearly there is
preferential alignment (as in numerical and experimental 3D flows
\cite{GeoVortExpNum}). To quantify the preferred  rate of strain
state, we display in Fig. \ref{fig:GeoVar}(b) the PDF of the
parameter
$s^*={-3\sqrt{6}\alpha\beta\gamma}/{(\alpha^2+\beta^2+\gamma^2)^{3/2}}$.
As in real flows \cite{GeoVortExpNum}, the PDF of $s^*$ is shifted
towards a peak at $s^*=1$ (axisymmetric expansion).

\begin{figure}[t]
\centerline{\epsfig{file=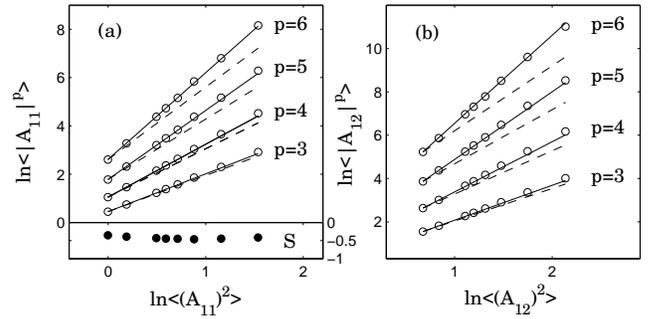,width=8.5cm}}
\caption{\label{fig:Anomalous} Relative scaling of velocity
gradient moments ($\circ$) in the (a) longitudinal direction and
(b) transverse direction for various orders $p$. Different points are for
various $\Gamma$ (from left to right $\Gamma = $
0.2, 0.15, 0.1, 0.09, 0.08, 0.07, 0.06, and 0.05).
In (a) the skewness coefficient of
longitudinal components $\mathcal S = \langle
A^3_{11}\rangle/\langle A_{11}^2\rangle^{3/2}$ is also shown
($\bullet$ using the right scale). Solid lines denote predictions
from multifractal scaling, dashed lines are Kolmogorov (1941)
non-anomalous scaling.}
\end{figure}

An important feature of small-scale turbulence is scaling of
higher-order moments with $\mathcal R_e$ \cite{Nel90}, i.e.
$\langle |A_{11}|^p\rangle \sim \mathcal R_e^{\mathcal F(p)}$.
Regular K41 scaling corresponds to $\mathcal F(p)=p/2$ \cite{K41}
while deviations indicate anomalous scaling. However, the simple
assumption to take the forcing term $\textbf{W}$ Gaussian and
delta correlated in time is expected to be  realistic at most for
a limited range of Reynolds numbers. Therefore, we present results
in terms of relative scaling which utilizes the above relation for
$p=2$ to obtain $\mathcal R_e\sim \langle A_{11}^2\rangle$ (using
$F(2)=1$ from  the condition of finite dissipation), and thus
$\langle |A_{11}|^p\rangle \sim \langle A_{11}^2\rangle^{\mathcal
F(p)}$. Shown in Fig. \ref{fig:Anomalous} are $p$-order moments of
$A_{11}$ and $A_{12}$, as functions of the second-order moments,
and varying parameter $\Gamma$.  Deviations from the dashed lines
(K41 case with slope $p/2$) are consistent with anomalous scaling.
Since PDFs of normalized $A_{11}$ and $A_{12}$ change with
$\Gamma$ or $\mathcal R_e$, their statistics cannot follow K41
scaling. The solid lines in Fig. \ref{fig:Anomalous} use the
multifractal formalism: $\mathcal F(p) = - \min_h
\{[p(h-1)+1-\mathcal D(h)]/(h+1)\}$ and $\mathcal D(h)$ is the
classical singularity spectrum \cite{Fri95}. The latter is used
here with a parabolic approximation $\mathcal D(h) =
1-{(h-c_1)^2}/{(2c_2)}$, with $c_1=1/3+3c_2/2$
\cite{Nel90,CheCas06}, and thus a single unknown parameter $c_2$
($c_2=\mu/9$, where  $\mu$ is the usual intermittency exponent).
The numerical results can thus be used to determine $c_2$ from the
model by fitting the slopes in  Fig. \ref{fig:Anomalous}. The
solid lines are for a parameter $c_2=0.025$ (or $\mu=9c_2 \sim
0.22$) for the longitudinal, and $c_2=0.040$ for the transverse
cases. These values are in excellent agreement with values found
from experiments and DNS \cite{Fri95,CheCas06}. The longitudinal
derivative skewness factor $\mathcal S$ shows characteristic
values near $-0.5$.

In conclusion, building on several prior works
\cite{Can91,GirPop90_1,ChePum99,JeoGir}, a new model has been
proposed for the anisotropic part of the pressure Hessian and the
viscous diffusion term entering in the Lagrangian evolution
equation for the velocity gradient tensor $\textbf{A}$. The system
predicts a variety of local, statistical, geometric and anomalous
scaling properties of 3-D turbulence.  Results are obtained within
a limited range of the parameter $\Gamma$, or Reynolds number
$\mathcal R_e$.  When tests are done with  $\Gamma$ below 0.05,
the PDFs of velocity increments, of $R$ and $Q$, and alignment
trends become less realistic. This is due possibly to the
limitations imposed by the assumption of Gaussian forcing. More
work is needed to extend the approach to arbitrarily high Reynolds
numbers, possibly by adding additional degrees of freedom to the
model or by modifying the type of forcing. Moreover, establishing
connections with the statistics of Lagrangian  structure functions
(velocity increments in time instead of the spatial  variations
described by $A_{ij}$) requires additional models to describe
jointly the Lagrangian evolution of velocity and velocity
gradients.

We gratefully acknowledge the Keck Foundation (LC) and the National Science
Foundation (CM) for financial support. We thank L. Biferale for
his very insightful comments, and Y. Li, Z. Xiao, B. Castaing,
G. Eyink, E. Vishniac, S. Chen and A. Szalay for fruitful discussions.


\end{document}